\documentclass[11pt,a4paper]{article}

\usepackage{amssymb,amsmath,amsthm}
\usepackage{color}
\usepackage{enumerate}
\usepackage[top=25mm, bottom=25mm, left=20mm, right=20mm]{geometry}
\usepackage{graphicx}
\usepackage{cite}

\DeclareMathOperator{\sech}{sech}

\title{\bf{Mathematics of nerve signals}}

\author{J\"{u}ri Engelbrecht, Kert Tamm, Tanel Peets  \\
	Department of Cybernetics, School of Science,	Tallinn University of Technology\\
	Akadeemia tee 21, Tallinn 12618, Estonia
	}

\begin{document}

\maketitle

\abstract{Mathematical models describing the signals propagating in nerve fibres are described. The emphasis is on the mathematical structures of governing equations while the extremely rich physiological aspects are here not analysed. Based on models of single waves, a joint coupled model is presented which is able to describe the action potential and the accompanying mechanical effects togehter with temperature changes within one system of partial differential equations. The whole signal is an ensemble which includes primary and secondary components. The primary components of a signal are the action potential itself and longitudinal mechanical waves in axoplasm and surrounding biomembrane. These components are characterized by corresponding velocities. The secondary components of a signal are derived from primary components and include transverse displacement of a biomembrane and the temperature change. These secondary components have no independent velocities in the presented model.}

\tableofcontents

%
%
%
%
%
%
%
%
%
%
%
%
%
%
%
%
%
%
%
%
%

\newpage
\bigskip

\section{Introduction}
\subsection{Nerve pulse and its accompanying effects.} Signal propagation in nervous systems is extremely important for all multicellular animals including human beings. Nerve signals control motion, behaviour and consciousness in many respects. The main structural element of a nervous system is the axon which ensures the signal propagation from a nerve cell to a nerve terminal. Axons are built as cylindrical biomembranes made of lipid bilayers and filled with fluid called axoplasm (intracellular fluid). This structure called also a nerve fibre is placed into a surrounding extracellular (intersticial) fluid. The electrical signal in the axoplasm is called action potential (AP). The contemporary understanding of axon physiology is described in many overviews \cite{Nelson2004,Clay2005,Debanne2011} etc. 
Beside the electrical signals which are supported by ionic currents through the biomembrane, there are also accompanying effects: deformation of the biomembrane, pressure in the axoplasm, temperature changes, etc. Usually most nerve fibres are covered by myelin sheath which permits the ion change between intracellular and extracellular fluids only in certain nodes (Ranvier nodes). The squid axon used in many experiments is, however, unmyelinated. Further on, we shall model the behaviour of unmyelinated axons.
The AP and the accompanying effects form actually an ensemble of waves. The general signal propagation can be described in a following way. The AP is generated by an electrical impulse with an amplitude above the threshold. The AP propagates in the axoplasm supported by ion currents through the biomembrane. The AP generates also a pressure wave (PW) in the axoplasm and a longitudinal wave (LW) in the biomembrane. The LW means a local longitudinal compression which changes the diameter of the biomembrane reflected by a transverse wave (TW).
These waves are characterized by the following values of their physical parameters:
\begin{enumerate}[(i)]
\item the AP has a resting potential in the axoplasm about -60 to -80 mV, AP increase (amplitude) is about 100--110 mV, its duration about 1 or less ms, its velocity depends on the diameter and varies from 10 to 110 m/s (estimates by many authors), the AP has an overshoot responsible for the refractory period; 
\item the PW has a value in the range of 1 to 10 mPa \cite{Terakawa1985}; 
\item the TW has a value about 1--2 nm \cite{Terakawa1985,Tasaki1988} etc.
\end{enumerate}
Based on measurements, the AP has an asymmetric shape with an overshoot with respect of the resting potential, the PW has a unipolar shape with a possible overshoot, the TW has a bipolar shape which corresponds to a unipolar LW. The temperature change during the passage of an AP is measured about $20-30 \mu$K \cite{Tasaki1988} for garfish but also much less for bullfrog \cite{Tasaki1992}. However, the experimental results under various conditions may differ from these estimations.

\subsection{Mathematical modelling.} Contemporary systems biology uses widely not only the experiments but also mathematical modelling and simulation \cite{Noble2002,Vendelin2007}.
The mathematical modelling is not the aim of per se, but a powerful tool for in silico simulations and understanding the possible mechanisms which govern the biological processes. The experimental verification of mathematical models is an important step in understanding the process in general as well as in details. A valuable side of a model is that it permits to make predictions about the process. In what follows, the mathematical models of signal propagation in nerve fibres are presented with a final aim to build up a joint mathematical model for describing the main coupled effects of relevant signals. The extremely rich morphology of axons and chemical processes are here not considered (see, for example \cite{Clay2005,Debanne2011}) and the attention is focused on basic mathematical models. As far as signals are treated as waves, the models use the basis of the classical wave equation which is one of the most important equations of mathematical physics \cite{Stewart2013}. 

The classical wave equation based on the Newton's $2^{nd}$ Law includes the acceleration and the force (the stress). Dealing with electrical signals, its counterpart is the telegraph equation which describes the voltage (and current) in a transmission line. In mathematical terms, both are the second order linear partial differential equations (for a lossless case, the telegraph equation is just the classical wave equation). However, for describing the electrical signals in nerves, the inductance is usually neglected resulting in a parabolic equation of the reaction-diffusion type. In order to deal with more complicated cases, these equations must be modified. This problem will be explained below for processes in nerve fibres. The general questions to be answered are: what is the use of such equations (models); should the equations be robust or detailed; what are the hints for reality obtained from these equations; what are the leading quantities from these equations, how the theoretical results could be verified, etc. One cannot forget the famous saying attributed to A. Einstein: Everything should be made as simple as possible but not simpler.

\section{Action potential}
\subsection{History.} The rich history of the problem is described in many overviews (see, for example, \cite{Bishop1956,Katz1966,Nelson2004}).
Already Luigi Galvani ($18^{th}$ century) has noticed the role electrical signals in muscle activities. Emil du Bois-Reymond (1818--1896) has determined the nerve signal as a ``negative variation". Julius Bernstein (1838--1917) has explained the signalling as a electro-chemical message caused by differences in ion concentration of intracellular and extracellular fluids. In contemporary terms, it was a start for the membrane theory. In $20^{th}$ century the physiological experiments were spiced with mathematical ideas and several mathematical models were proposed. This tendency is going on. One should stress the importance of the Nernst potential (Walter H Nernst, 1864--1941) in calculating the ion concentration across the biomembrane which determines the ion currents. A concise history of neuroscience is given by Alwyn Scott \cite{scott2002} who also stresses the role of mathematical studies in neuroscience. It is worth to notice that the studies on nerve signalling have been in focus for a long time and over 1906--2000 altogether 9 Nobel Prizes in physiology and medicine have been awarded to pioneering results in these studies.
Further the models derived by Hodgkin and Huxley, Bonhoeffer, FitzHugh, Nagumo,  etc are briefly described in mathematical terms.

\subsection{Hodgkin-Huxley model}

This is nowadays considered as the classical model for the nerve signal propagation. Hodgkin  and Huxley \cite{Hodgkin1952} have used the telegraph equation neglecting the inductivity. Further we use the presentation of Scott \cite{scott1999} for describing the Hodgkin-Huxley (HH) model (see also \cite{Engelbrecht1991}). From the telegraph equation a parabolic equation in space and time coordinates is obtained
\begin{equation}\label{2.1}
\frac{\partial^{2} V} {\partial x^{2}} - RC_{A} \frac{\partial V} {\partial t} = Rj,
\end{equation}
where $V$ is the voltage, $R$ is the longitudinal resistance, $C_A$ is the membrane capacitance and $j$ is the total ion current; $R, C_A$ and $j$ are all given per unit length. Ion current according to Hodgkin-Huxley \cite{Hodgkin1952} can be calculated by
\begin{equation}
\label{22}
j = \pi d J,
\end{equation}
\begin{equation}
\label{23}
J_i=G_K n^{4}(V-V_K) + G_{Na} m^{3}h (V-V_{Na}) + G_L (V-V_{L})+C_m \frac{\partial V} {\partial t},
\end{equation}
where $d$ is the axon diameter, $G_K$ and $G_{Na}$ are, respectively, the potassium and sodium conductance, $G_L$ is a leakage conductance, $C_m$ is the membrane capacitance and $V_K, V_{Na}, V_L$ are the corresponding equilibrium potentials.

The phenomenological variables $n, m$ and $h$ govern the ``turning on" and ``turning off" the membrane conductance. Namely the potassium conductance is turned on by $n$ and sodium conductance by $m$ while the sodium conductance is turned off by $h$. In terms of continuum mechanics, these variables can also be called internal variables \cite{Maugin1994}. Variables $n,m,h$ are governed by relaxation equations
\begin{equation}\label{2.4a}
dn/dt = \alpha_n (1-n) - \beta_n n,
\end{equation}
\begin{equation}\label{2.4b}
dm/dt = \alpha_m (1-m) - \beta_m m,
\end{equation}
\begin{equation}\label{2.4c}
dh/dt = \alpha_h (1-h) - \beta_h h,
\end{equation}
where the coefficients are determined by  
\begin{equation}\label{2.5a}
\alpha_n = \frac{0.01(10-V)}{\mathrm{exp}[(10-V)/10]-1},
\end{equation}
\begin{equation}\label{2.5b}
\beta_n = 0.125{\mathrm{exp}}(-V/80),
\end{equation}
\begin{equation}\label{2.5c}
\alpha_m = \frac{0.1(25-V)}{\mathrm{exp}[(25-V)/10]-1},
\end{equation}
\begin{equation}\label{2.5d}
\beta_m = 4{\mathrm{exp}}(-V/18),
\end{equation}
\begin{equation}\label{2.5e}
\alpha_h = 0.07{\mathrm{exp}}(-V/20),
\end{equation}
\begin{equation}\label{2.5f}
\beta_h = \frac{1}{\mathrm{exp}[(30-V)/10]+1}.
\end{equation}

These coefficients are in units of $(msec)^{-1}$ while voltage $V$ is in millivolts.

The celebrated HH model has been tested by many experiments. Several modifications have been proposed to modify the ion current $j$ (see, for example, \cite{Courtemanche1998}). However, the HH model describes the AP like an electrical signal with ion currents and does not take into account other accompanying effects \cite{Appali2010}.

\subsection{Bonhoeffer-van der Pol model}

The  HH model involves many variables and many physical parameters which depend on the morphology of nerve fibres. This is why the attention was also to deriving more simple models which could describe the process with sufficient accuracy. One such an attempt is described by FitzHugh \cite{FitzHugh1961}. He has used the ideas of Bonhoeffer \cite{Bonhoeffer1948} and van der Pol \cite{vanderpol1926} for deriving a model of excitable-oscillatory system. Here and below the notations from original papers are used. The starting point is the van der Pol equation
\begin{equation}\label{2.6}
\ddot{x} + c(x^{2}-1)\dot{x}+x=0,
\end{equation}
where $x$ is an oscillating quantity (amplitude) and $c$ is a positive constant. The dots denote differentiation with respect to time $t$. By using Li\'{e}nard's transformation
\begin{equation}\label{2.7}
y= \dot{x}/c + x^{3}/3-x,
\end{equation}
the system of the first order equations is obtained:
\begin{equation}\label{2.8a}
\dot{x}=c (y+x-x^{3}/3),
\end{equation}
\begin{equation}\label{2.8b}
\dot{y}= - x/c.
\end{equation}

The Bonhoeffer-van der Pol (BVP) model enlarges the system~\eqref{2.8a}, \eqref{2.8b} to
\begin{equation}\label{2.9a}
\dot{x}=c (y+x-x^{3}/3+ z),
\end{equation}
\begin{equation}\label{2.9b}
\dot{y}= - (x-a+by)/c,
\end{equation}
where
\begin{equation}\label{2.10}
1-2b/3<a<1,\,\,\,0<b<1,\,\,\,b<c^{2}.
\end{equation}

Here $a$ and $b$ are constants,  $z$ is the stimulus intensity representing the membrane current.

It is noted that $x$ is related to membrane potential and excitability and $y$ -- to accommodation and refractoriness. The phase-plane analysis is used for demonstrating the threshold and all-or-non phenomena together with the refractory part of solution. Numerical solution using $a=0.7, b=0.8, c = 3$ and various values of $z$ are obtained (\cite{FitzHugh1961}, Fig. 2).

In terms of variables, the variables $x,y,z$ in the BVP model correspond to variables $V$ and $m$, $h$ and $n, J_i$ in the HH model, respectively. However, the HH model describes the propagating  wave (Eq.~\eqref{2.1}  is an PDE) while the BVP model describes the standing profile (Eqs. (2.9) are ODE's).

\subsection{FitzHugh-Nagumo model}

In original notations the FitzHugh-Nagumo (FHN) model is written like the BVP model (2.9) in the form of a system \cite{Nagumo1962}:
\begin{equation}\label{2.11a}
h\frac{\partial^{2} u} {\partial s^{2}} = \frac{1}{c} \frac{\partial u} {\partial t} - w -\left ( u-\frac{u^{3}}{3} \right),
\end{equation}
\begin{equation}\label{2.11b}
c\frac{\partial w} {\partial t} + bw = a-u,
\end{equation}
where $u$ is the voltage and $w$ - the recovery current. The constants $h,c,b,a$ are positive satisfying
\begin{equation}\label{2.12}
1>b>0, \,\,\, c^{2}>b, \,\,\,1>a>1-\frac{2}{3}b.
\end{equation}

As noted by Nagumo et al \cite{Nagumo1962}, system (2.11) is the distributed BVP model (2.9) which is an ODE. The possible governing equation can be derived from (2.11)
\begin{equation}\label{2.13}
ch \frac{\partial^{2} u} {\partial t s^{2}} = \frac{\partial^{2} u} {\partial t^{2}}-  c\left(1-u^{2}\right) \frac{\partial u} {\partial t} + u -a,
\end{equation}
where for simplicity $b=0.$

Once the BVP model describes a standing (stationary) wave, then the FHN model (2.13) results in
\begin{equation}\label{2.14}
\beta y^{'''} - y^{''} - \mu \left(1-y+\varepsilon y^{2}\right)y' - y = 0,
\end{equation}
where $y(\tau) \equiv z(x,t), \,\,\, \tau = t - x/\theta$ and $y'=dy/d\tau, \,\,\,\beta=\theta^{-2}>0$ while $z$ is nondimensional voltage (note that originally $\xi\equiv y$).

A modification of Eqs (2.11) changes the polynomical $u-1/3 u^{3}$ to its full form \cite{Neu1997,Bountis2000}.
Then Eqs (2.11) can be represented in the form  of two coupled equations \cite{Engelbrecht2018} using the dimensionless variables $z$  and $j$:
\begin{equation}\label{2.15a}
\frac{\partial z} {\partial t} = z (z-a)(1-z)-j+D \frac{\partial^{2} z} {\partial x^{2}},
\end{equation}
\begin{equation}\label{2.15b}
\frac{\partial j} {\partial t}  = \varepsilon (-j+bz),
\end{equation}
where $D,a,b,\varepsilon$ are constants.

\subsection{Evolution equations.}

From wave mechanics it is known that the classical wave equations describes two waves -- one propagating to the right, another -- to the left. Under certain conditions these waves can be separated and in this case the result is an evolution equation which describes one wave -- either propagating to the right or to the left. In mathematical terms the leading derivative is then of the first order. The details of such derivation by the reductive perturbation method are described in several monographs \cite{taniuti1983,Engelbrecht1983}

The methods governing the propagation  of the AP are as a rule derived from the telegraph equation (see Eq. (2.1), for example) where the inductance is neglected. The result is a parabolic equation (2.1) which due to the existence of the ion current $j$ leads to the propagating wave. Returning to the initial telegraph equation, the governing equation is hyperbolic \cite{Lieberstein1967}. Then it is possible to apply the reductive perturbation method and derive the corresponding evolution equation.

The full telegraph equation with the ion current added is \cite{Engelbrecht1981}
\begin{equation}\label{2.16}
\frac{\partial^{2} V} {\partial x^{2}}-LC_A \frac{\partial^{2} V} {\partial t^{2}} -  RC_A \frac{\partial V} {\partial t} = \frac{2}{a}R_j + \frac{2}{a}L\frac{\partial j}{\partial t} ,
\end{equation}
where in addition to notations in Eq.~\eqref{2.1}  $L$ denotes the inductance and $a$ -- the axon radius. The phase velocity $c_0$ from Eq (2.16) is
\begin{equation}\label{2.17}
c_0= (1/LC_A)^{1/2},
\end{equation}
which is not the final velocity of the pulse but serves as a basis for applying the reductive perturbation method. The evolution equation yields in the following form \cite{Engelbrecht1981,Engelbrecht1991}: 
\begin{equation}\label{2.18a}
\frac{\partial^{2} z} {\partial \xi \partial x} + f(z) \frac{\partial z} {\partial \xi}+g(z) =0,
\end{equation}
\begin{equation}\label{2.18b}
f(z) = b_0 + b_1z + b_2z^{2},
\end{equation}
\begin{equation}\label{2.18c}
g(z) = b_{00} z,
\end{equation}
where $z$ is a scaled voltage, $\xi=c_0t-x$ and $b_0, b_1, b_2, b_{00}$  are constants. Here the ion current is used after Nagumo et al.~\cite{Nagumo1962}
\begin{equation}\label{2.19a}
j = k_1u + k_3u^{3} + w,
\end{equation}
\begin{equation}\label{2.19b}
\frac{\partial w}{\partial t} = q_0 (u+q_1),
\end{equation}
where $k_1, k_2, k_3, q_0, q_1$ are constants. The roots $z_1, z_2$ of the polynomial ~\eqref{2.18b} are both positive.

It is possible to derive the equation governing a stationary profile like in the case of the BVP model.

Suppose $\tau=x+\theta \xi$, where $\theta$ is the pseudovelocity. Then~\eqref{2.18a} yields
\begin{equation}\label{2.20}
z^{\prime\prime} + f(z) z' + \theta^{-1}\xi = 0,
\end{equation}
where $z' = dz / d\tau$. Equation~\eqref{2.20}  is a Li\'{e}nard-type equation which governs the stationary pulse (cf.~Eq.~\eqref{2.14}).

The details about the solutions of Eq. (2.18) and Eq~\eqref{2.20}  are described by Engelbrecht \cite{Engelbrecht1991}.

\section{Biomembrane}

A cylindrical biomembrane forms a wall of a nerve fibre separating intra- and extra-cellular fluids. It is built by a lipid bi-layer which in terms of mechanics can be treated as a microstructured tube. Such a biomembrane can be deformed and according to present understanding takes  also part in signal formation and propagation. Leaving aside the physiological aspects of the biomembrane (see, for example, \cite{Mueller2014}) we present here the mathematical model for describing its longitudinal deformation.

Such a mathematical model for a longitudinal wave (LW) is proposed by Heimburg and Jackson \cite{Heimburg2005}. It is assumed that the compressibility of a biomembrane has an impact on the velocity $c$:
\begin{equation}\label{3.1}
c^{2} =c_0^{2}+ p u + q u^{2},
\end{equation}
where $c_0$ is the initial sound velocity, $u= \triangle \rho_A, \rho_A$ is the density and $p, q$ are constants. Then from the traditional 1D wave equation the governing equation for the LW is derived:
\begin{equation}\label{3.2}
\frac{\partial^{2} u} {\partial t^{2}} = \frac{\partial} {\partial x} \left[\left(c_0^{2}+pu+qu^{2}\right) \frac{\partial u} {\partial x} \right] -h_1 \frac{\partial^{4} u} {\partial x^{4}},
\end{equation}
where the last term with constant $h_1$ is added in order to model dispersive effects. Later Engelbrecht et al.~\cite{EngelbrechtTammPeets2014} have shown  that in order to model better the microstructured effects in a biomembrane, an additional dispersion term must be added which reflects the inertial properties. In this case the governing equation takes the form
\begin{equation}\label{3.3}
\frac{\partial^{2} u} {\partial t^{2}} = \frac{\partial} {\partial x} \left[\left(c_0^{2}+pu+qu^{2}\right) \frac{\partial u} {\partial x} \right ] -h_1 \frac{\partial^{4} u} {\partial x^{4}} + h_2 \frac{\partial^{4} u} {\partial x^{2}\partial t^{2}},
\end{equation}
where now two dispersive constants $h_1$ and $h_2$ and respective derivatives govern the influence of dispersion. Equation~\eqref{3.3}  is a Boussinesq-type equation with the amplitude-dependent nonlinearities.

\section{Axoplasm}

The axoplasm within a nerve fibre can be modelled like a viscous fluid \cite{Gilbert1975}. It means that a pressure wave (PW) can be described by Navier-Stokes equations. In the 1D setting such model in terms of longitudinal velocity $v_x$ is \cite{Tritton1988}
\begin{equation}\label{4.1}
\rho \left (\frac{\partial v_x}  {\partial t} + v_x \frac{\partial v_x} {\partial x} \right ) + \frac{\partial \bar{p}} {\partial x}  - \mu_1 \frac{\partial^{2} v_x} {\partial x^{2}} = F,
\end{equation}
where $\rho$ is the density, $\bar{p}$ is the pressure, $\mu_1$ is the viscosity parameter and $F$ - the body force.

In the 2D setting for waves on the fluid surrounded by a cylindrical tube, the governing equations are \cite{Lin1956}: 
\begin{equation}\label{4.2a}
\frac{\partial^{2} \bar{p}}  {\partial t^{2}} = c_f^{2}\left ( \frac{\partial^{2} \bar{p}} {\partial x^{2}}+\frac {\partial^{2} \bar{p}} {\partial r^{2}}
  + \frac{1}{r}    \frac{\partial \bar{p}} {\partial r} \right ),
\end{equation}
\begin{equation}\label{4.2b}
\rho \frac{\partial^{2} U_x}  {\partial t^{2}} +  \frac{\partial \bar{p}} {\partial x} =0,
\end{equation}
\begin{equation}\label{4.2c}
\rho \frac{\partial^{2} U_r}  {\partial t^{2}} +  \frac{\partial \bar{p}} {\partial r} =0,
\end{equation}
where $x,r$ are cylindrical coordinates, $U_x$ and $U_r$ are longitudinal and transverse displacements, respectively and $c_f$ is the velocity.

While the amplitude of the pressure wave (PW) is very small \cite{Terakawa1985} then it is possible also to model PW by the wave equation
\begin{equation}\label{4.3}
\frac{\partial^{2} \bar{p}}  {\partial t^{2}} = c_f^{2}  \frac{\partial^{2} \bar{p}} {\partial x^{2}}- \mu \frac{\partial \bar{p}} {\partial t},
\end{equation}
where $\mu$ is the viscosity parameter.

\section{Temperature changes}
There are several ideas proposed in earlier studies on thermal changes caused by propagating APs.
It has been stated that the energy of the membrane capacitor $E_c$ is \cite{Ritchie1985}
\begin{equation}
\label{Eceq}
E_c = \frac{1}{2} C_m Z^2,
\end{equation}
where $C_m$ is a capacitance and $Z$ is the amplitude of the AP (here and further in Section 5 dimensionless variables are used). Noting that heat energy $Q \approx E_c$, it can be deduced that $Q$ should be proportional to $Z^2$. The standard Fourier law (the thermal conductivity equation) in its simplest 1D form is
\begin{equation}
\label{FourierL}
Q = -k \Theta_X,
\end{equation}
where $Q$ is the heat energy, $k$ is the thermal conductivity and $\Theta$ is the temperature. Note that heat energy is proportional to the negative temperature gradient. Combining \eqref{Eceq} and \eqref{FourierL}, we obtain 
\begin{equation}
\label{ThetaEq}
\Theta \propto  \frac{C_m}{2 k} \int Z^2 dX.
\end{equation}
Further we follow experimental results by Abbott et al.~\cite{Abbott1958} which demonstrate that the heat increase at the surface of the fibre is positive. 
It has also been argued that either $Z \propto \Theta$ or $Z \propto \Theta_T$ depending on physical properties \cite{Tasaki1992}. The potential $Z$ of the AP or its square $Z^2$ might be not the only sources. 
For example, in \cite{Heimburg2008,Schneider2018} it is argued in favour of the idea that experimentally observed temperature changes might be the result of the propagating mechanical wave in the lipid bi-layer. As there seems to be no clear consensus which quantities associated with the nerve pulse propagation are the sources of the thermal energy, we will consider here several possibilities for the coupling of the waves in the ensemble to the additional model equation for the temperature. 

The idea is to cast these ideas into a mathematical form.
As far as temperature is a function of space and time
we opt to use the classical heat equation  which is a parabolic PDE describing the distribution of heat (or variation in temperature) in a given region over time. 
It is straightforward \cite{CarslawJaeger1959} to derive the heat equation in terms of temperature from the Fourier's law by considering the conservation of energy 
\begin{equation}
\label{HHeq}
\Theta_{T} = \alpha \Theta_{XX},
\end{equation}
where $\alpha$ is the thermal diffusivity. In our case, a possible source term must be added to Eq.~\eqref{HHeq}
\begin{equation}
\label{Heq}
\Theta_{T} = \alpha \Theta_{XX} + F_3(Z,J,U).
\end{equation}
The source term $F_3$ permits to account for different assumptions proposed to describe the temperature generation and consumption.

An overview on mathematical foundation of heat production and temperature changes in fibres is presented by Tamm et al.~\cite{Tamm2018b}.

\section{Modelling of an ensemble of waves}
\subsection{General ideas of modelling.} 
The idea to describe all the effects of signal propagation in nerve fibres is quite natural. Such an idea was stated already by Hodgkin \cite{Hodgkin1964} and more recently by Andersen et al.~\cite{Andersen2009}. It is a real challenge to build up a mathematical model involving the AP, PW, LW and TW into one system. Above (see Sections~2--5) the mathematical models governing the single waves AP, PW, LW  were presented. In order to unite these equations into one system involving the coupling effects, the crucial part must be added -- the coupling forces. 

The mathematical models described in Sections 2 to 4 were based on wave equations modified in order to grasp additional accompanying effects. In case of the AP, the HH and FHN models represent the reaction-diffusion type equations which were derived from the wave equation neglecting the inductivity. The AP is generated by an initial input which must exceed a certain threshold. 

In case of the PW and the LW, the basic models are clearly the hyperbolic wave equations. In order to be included into a joint model, one should understand their behavior under the forcing.

There are several possibilities to analyse the forced wave equation
\begin{equation}
\label{Fweq}
\frac{\partial^2 u}{\partial t^2} - c^2 \frac{\partial^2 u}{\partial x^2} = f(x,t),
\end{equation}
where $f(x,t)$ is a smooth function. It is possible to solve eq.~\eqref{Fweq} analytically in series \cite{Dai2008} or to use the Green's function \cite{Graff1991}. In the latter case, eq.~\eqref{Fweq} is first solved for a unit load
\begin{equation}
\label{Fweqdelta}
f(x,t) = \delta (x-\xi) \delta(t-\tau),
\end{equation}
where $\delta$ is the delta function and for an arbitrary $f(x,t)$ the solution is
\begin{equation}
\label{Fweqsolution}
u(x,t) = \int^{t}_{0} d \tau \int_{-\infty}^{+\infty} G \left(x, \frac{t}{\xi}, t \right) f (\xi, \tau) d \xi,
\end{equation}
where $G(x,t,\xi)$ takes into account the initial conditions \cite{Graff1991}. 

Further on our interest is to the pulse-type or bi-polar-type forcing and to numerical solutions using the pseudo-spectral method (PSM). The tests have shown that the bi-polar forcing to a linear wave equation leads to energetically more balanced solution. 

There are several proposals to build up a joint system \cite{Jerusalem2014,Hady2015}. To the best knowledge of authors, the ideas to compose the system of governing differential equation are proposed by Engelbrecht et al.~\cite{Engelbrecht2018,Engelbrecht2018c,Engelbrecht2018d}. Further we represent their model based on results described in Sections 2--5.

\begin{figure}[h] \label{Fig1}
\includegraphics[width=0.43\textwidth]{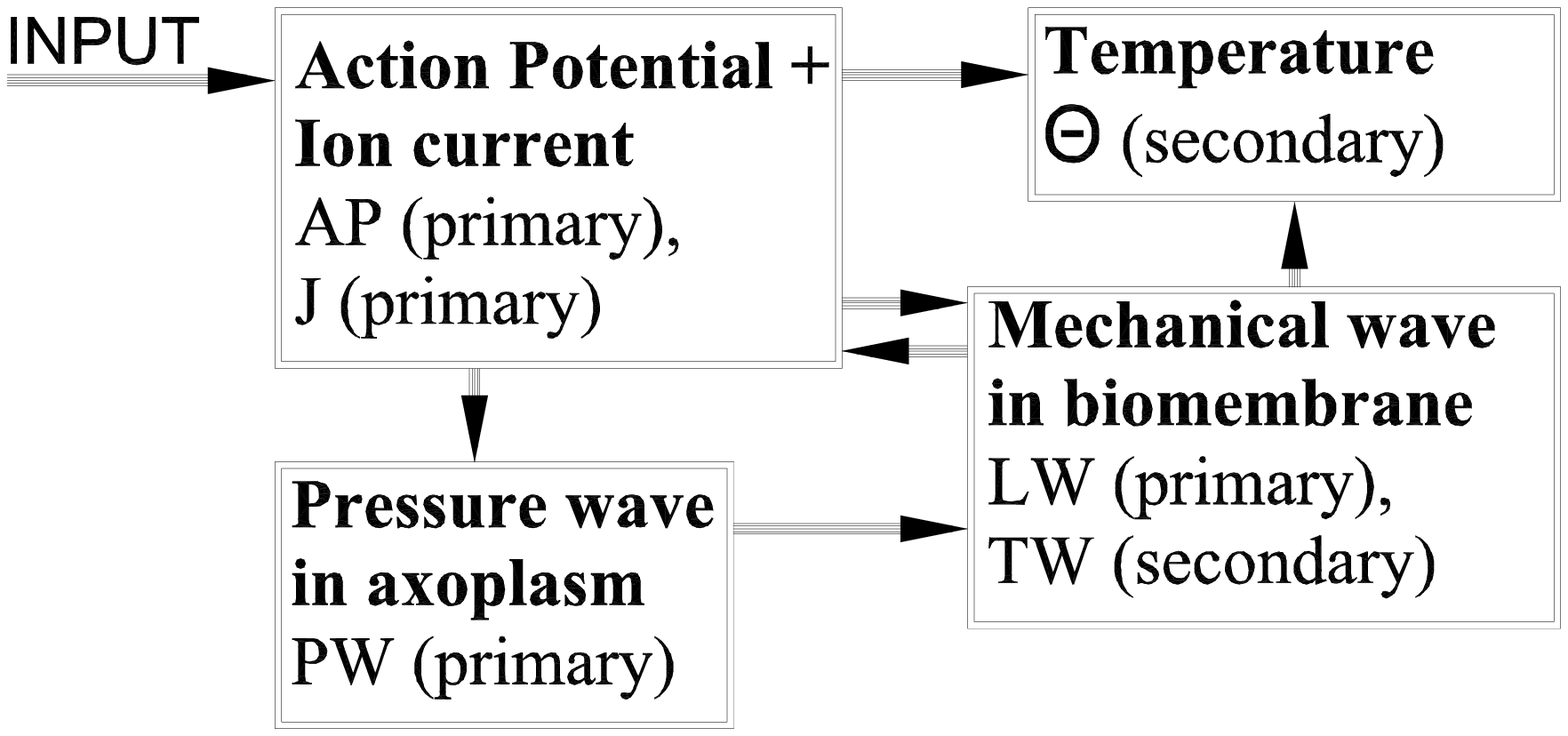}
\includegraphics[width=0.56\textwidth]{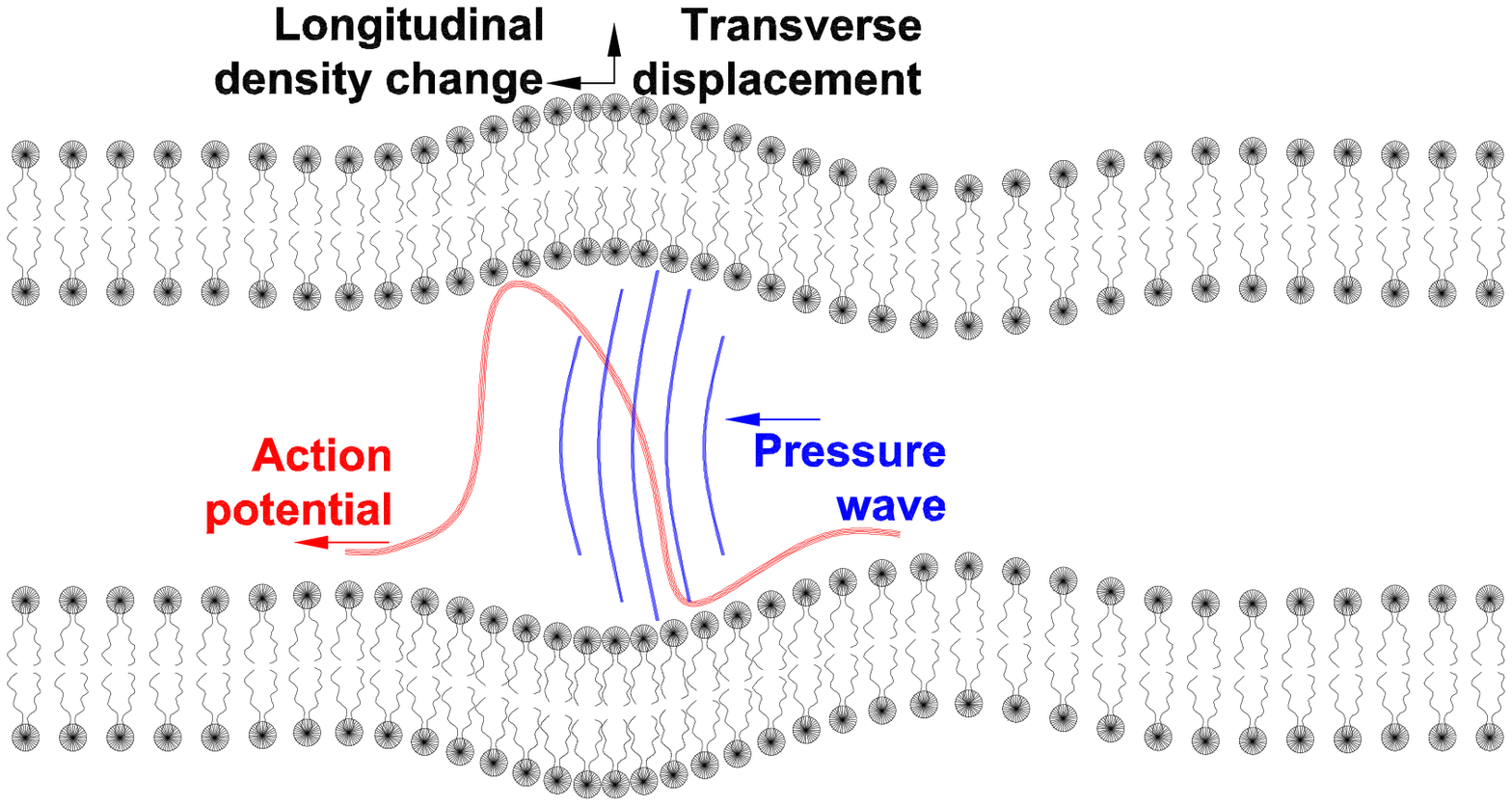}
\caption{The block diagram of the possible mathematical model for the ensemble of waves in the nerve fiber (left) and an artistic sketch of wave ensemble propagation in the axon (right). Here AP is the action potential, J is the ion current, LW is the longitudinal wave in the biomembrane, PW is the pressure wave in axoplasm, $\Theta$ is the temperature and TW is the transverse wave in the biomembrane.}
\end{figure}

In general terms, the wave ensemble in a nerve fibre is composed by primary components: the AP, the PW, and the LW. These waves when considered separately possess finite velocities but in the ensemble, the velocities are synchronized. The synchronization demonstrated in experiments is a process which needs further analysis. The secondary (i.e., derived) components are the TW and temperature change TMP.  These components have no independent velocities and cannot emerge without primary components. In this way, the main physical properties of a fibre are taken into account by the primary components. The block-scheme of signal generation with primary and secondary components of an ensemble of waves is shown in Fig.~\ref{Fig1}.

\subsection{Engelbrecht -- Tamm -- Peets model}

We follow now the main ideology of  Engelbrecht et al.~\cite{Engelbrecht2018c}. The process is initiated by an electrical impulse $z(t)$
\begin{equation}\label{5.1}
 z \bigm|_{t=0} = f(x),
\end{equation}
where $z$ is above the thereshold level. The AP is described by Eq. (2.15) rewritten in the form
\begin{equation}\label{5.2a}
z_{t} = z\bigl ( z-(a_1+b_1) \bigr) (1-z) - j + Dz_{xx},
\end{equation}
\begin{equation}\label{5.2b}
jt = \varepsilon\bigl ( -j+(a_2+b_2) z\bigr),
\end{equation}
where parameters  $a_1, a_2$ control the `electrical' activation and parameters $b_1, b_2$ control the `mechanical' activation, while $\varepsilon, D$ are additional constants, indices $x$ and $t$ here and further denote differentiation.

The PW in the axoplasm is governed by Eq.~\eqref{4.3}  with a driving force
\begin{equation}\label{5.3}
\bar{p}_{tt} = c_f^{2}\bar{p}_{xx} - \mu \bar{p}_t + f_1 (z,j),
\end{equation}
where $f_1(z,j)$ is a force from the AP.

The LW in the biomembrane is governed by Eq.~\eqref{3.3}
\begin{equation}\label{5.4}
u_{tt} = \Bigl [\left(c_0^{2}+pu+qu^{2}\right)u_{x} \Bigr ]_x - h_1u_{xxxx} + h_2 u_{xxtt} + f_2(j,\bar{p}),
\end{equation}
where $f_2(j,\bar{p})$ is a force exerted by the processes (AP and PW) in the axoplasm.

The transverse wave TW following the ideas from the mechanics of rods \cite{Porubov2003} is governed by
\begin{equation}\label{5.5}
w=-kru_{x},
\end{equation}
where $r$ is the radius of the fibre and $k$ is a constant (in the theory of rods the Poisson ratio).

The crucial problem is to assume the mechanism of coupling forces. The main hypothesis \cite{Engelbrecht2018c,Engelbrecht2018d} is the following: the mechanical waves in axoplasm and surrounding biomembrane are generated due to changes in electrical signals (the AP and ion current). The seconding hypothesis is: the changes in the pressure wave PW may also influence the waves in biomembrane. As far as changes are related to either space or time derivatives, it is assumed that the leading terms in coupling forces are in the form of $z_x, z_t, j_x, j_t, \bar{p}_{x}, \bar{p}_{t} $.

The final form of a joint model in dimensionless variables and coordinates is \cite{Engelbrecht2018c,Engelbrecht2018d} the following. The primary components:
\begin{equation}\label{5.6a}
Z_T= D_1 Z_{XX}-J+Z \bigl [Z - \left(A_1+B_1\right) - Z^{2}+(A_1+B)Z\bigr ],
\end{equation}
\begin{equation}\label{5.6b}
J_T= \varepsilon_1 \bigl [\left(A_2+B_2\right)  Z- J\bigr ],
\end{equation}
\begin{equation}\label{5.6c}
\begin{split}
U_{TT}= & c^{2}U_{XX} + NUU_{XX} + MU^{2}U_{XX} + MU^{2}U_{XX} + NU_{X}^{2} + \\
& + 2MUU_X^{2} - H_1U_{XXXX}+H_{2}U_{XXTT} + F_1(P,J,Z),
\end{split}
\end{equation}
\begin{equation}\label{5.6d}
P_{TT}= c_f^{2}P_{XX} -\mu P_T + F_2(Z_J),
\end{equation}
\begin{equation}\label{5.6e}
F_1= \gamma_1 P_T + \gamma_2 J_T - \gamma_3 Z_T,
\end{equation}
\begin{equation}\label{5.6f}
F_2= \gamma_1 Z_X + \gamma_2 J_T + \gamma_3 Z_T.
\end{equation}
The secondary effects: the TW is calculated as 
\begin{equation}
\label{TWeq}
W = K U_X,
\end{equation}
and temperature $\Theta$ is governed by 
\begin{equation}
\label{ThetaEQ}
\Theta_{T} = \alpha \Theta_{XX} + F_3\left(Z,J,U\right).
\end{equation}
Several coupling forces $F_3$ are used:
\begin{equation} \label{Fkolm}
F_3 = \tau_1 Z; \quad F_3 = \tau_2 Z^2; \quad F_3 = \tau_3 Z_T + \tau_4 J_T.
\end{equation}
An overview of distinguishing the primary and secondary components is presented by Engelbrecht et al.~\cite{Engelbrecht2018arXiv}. 

Here capital letters $Z, J, U, P, W, \Theta$ indicate variables  while $X, T$ are coordinates and other notations are parameters. The system of Eqs \eqref{5.6a}--\eqref{5.6f} is solved for
\begin{equation}\label{5.7}
Z|_{T=0} = F(X),
\end{equation}
where $F(X)$ is a pulse-like excitation.

\section{Final remarks}

The system \eqref{5.6a} to \eqref{5.6f} is solved as an initial value problem for an initial input~\eqref{5.7}. Equations~\eqref{5.6a} and ~\eqref{5.6b} describe the AP, Eq~\eqref{5.6c} describes the LW while Eq~\eqref{5.6d} describes the PW. The system is based on the 1D wave equations: the AP is described by a reaction-diffusion type equation which was derived from a wave equation and where the additional current has an important role, the LW and PW are described by modified wave equations. In mathematical terms it means that the process includes waves propagating to the right and to the left. It might be a challenge to use the evolution equation (one-wave equations) for describing the process. A possible evolution equation for the AP is Eq (2.18), the corresponding model for the PW is easily derived from Eq~\eqref{4.3}. The problem is with the corresponding model for the LW because the original model Eq~\eqref{3.3}  involves amplitude-dependent nonlinearities together with double dispersion and the usual reductive perturbation method cannot be directly applied. The possible simplifications might not be justified physically (see \cite{FongangAchu2018}).
Turning to the physical side the following must be stressed:
\begin{enumerate}[(i)]
\item the proposed model is a robust one but describes qualitatively correctly the profiles of essential signal components in a nerve fibre;
\item the AP is described by the FHN model which takes only one ion current into account (it is assumed to be similar to the sodium current \cite{Jacquir2008});
\item the model at this step can be taken as a proof of concept and the possible next step to modify is to use the HH model instead of the FHN model;
\item the important novel hypothesis for constructing the joint model is related to the coupling forces between the single components of the full signal; 
\item it is assumed that the coupling forces depend on changes of coupled signals (pulses), not on their amplitudes; 
\item by generating the AP from an initial input, the other waves are generated due to the coupling forces;
\item the properties of the TW which has been measured in several experiments depend on the mechanical properties of the biomembrane through the LW; 
\end{enumerate}

For the analysis of system \eqref{5.6a} to \eqref{Fkolm} or the processes of single AP, PW and LW various methods are used. The full system \eqref{5.6a} to \eqref{5.6f} needs numerical methods for the analysis. The results of such numerical simulations are presented by Engelbrecht et al.~\cite{Engelbrecht2018,Engelbrecht2018c,Engelbrecht2018d} and the corresponding numerical scheme is briefly described also in Appendix. There exists a huge depository of results related to the analysis of single processes of propagation of the AP, PW and LW. Note for example the overviews and research papers by \cite{FitzHugh1961,Nagumo1962,Hodgkin1964,Katz1966,Engelbrecht1991,scott2002,Heimburg2005}
-- just a few among many. In the studies of stationary profiles within simplified models of the FHN-type, the phase-plane analysis is used \cite{FitzHugh1961,Nagumo1962} etc. For the LW, the model equation  permits a closed stationary soliton-type solution \cite{Heimburg2005}, see also \cite{Peets2018}.
From a viewpoint of general wave mechanics, the final result described by a system of equations is an ensemble of waves demonstrating the complexity of the process. From a viewpoint of mechanics, the nerve fibre itself is a structure which involves microstructured medium (the biomembrane), localized solitary waves and possible internal variables (ion currents). The process is highly nonlinear which is characteristic to complex processes \cite{Engelbrecht2010}. As stressed by Coveney and Fowler \cite{Coveney2005}, coupled models across several processes could provide also the route for calculating many unknown parameters. That explains why systems biology turns so much attention to the modelling of biological complexity involving molecular and continuum approaches. This is also the case of nerve signals where the molecular structure of the biomembrane affects the process and the conjectured coupling forces need further studies and quantization.
Although much is known about the fascinating process of nerve signalling, the full picture needs further concerted efforts of experimentalists and theoreticians. A recent overview \cite{Drukarch2018a} reflects the contemporary insights in this field that could lead to ``the formulation of a more extensive and complete conception of the nerve impulse". The questions pointed out in Section 1.2 are not answered here but we hope that the framework of modelling ideas described above has explained the basis of the problem and will help to formulate such a complete conception together with finding the answers to these questions.

\noindent \section*{Appendix -- the numerical scheme}
\addcontentsline{toc}{section}{Appendix -- the numerical scheme}
This description was  earlier deposited in arXiv as a part of a research paper \\ (arXiv:1802.07014v[physics.bio.ph] 21 Feb 2018)
\section*{The system of partial differential equations to be solved numerically}
\def\theequation{A\arabic{equation}}
\setcounter{equation}{0}

As noted above, the pseudospectral method (PSM) (see \cite{Fornberg1998,salup2009}) is used to solve the system of dimensionless model equations:
\begin{equation}
\begin{split}
& Z_{T} = D Z_{XX} + Z \left[ Z - \left( A_1 + B_1 \right) - Z^2 + \left( A_1 + B_1 \right) Z \right] - J, \\
& J_{T} = \varepsilon \left[ \left( A_2 + B_2 \right) Z - J \right],\\
& U_{TT} = c^2 U_{XX} + N U U_{XX} + M U^2 U_{XX} + N U_{X}^{2} + 2 M U U_{X}^{2} -\\
& \quad \quad \quad H_1 U_{XXXX} + H_2 U_{XXTT} + \gamma_1 P_T + \gamma_2 J_T,\\
& P_{TT} = c_{f}^{2}P_{XX} + \eta_1 Z_X + \eta_2 J_T - \mu P_T,\\
& \Theta_{T} = \alpha \Theta_{XX} + F_3\left(Z,J,U\right).
\end{split}
\label{EQS}
\end{equation}
The notation used is already given above but is repeated here for the sake of completeness. Here $Z$ is the action potential, $J$ is the recovery current, $A_i, B_i$ are the `electrical' and `mechanical' activation coefficients, $D, \varepsilon$ are coefficients, $U=\Delta \rho$ is the longitudinal density change in lipid layer, $c$ is velocity of unperturbed state in lipid bi-layer, $M, N$ are the nonlinear coefficients, $H_1, H_2$ are the dispersion coefficients and $\gamma_1, \gamma_2$ are the coupling coefficients for the mechanical wave, $P$ is the pressure, $c_{f}$ is the characteristic velocity in the fluid, $\eta_1, \eta_2$ are the coupling coefficients for the pressure wave and $\mu$ is the (viscous) dampening coefficient. Notation $\Theta$ represents (local) temperature and $\alpha$ is a coefficient characterizing the temperature diffusion within the environment. `Mechanical' activation coefficients in the action potential and ion current expressions are connected to the improved Heimburg-Jackson part of the model as $B_1 = - \beta_1 U$ and $B_2 = - \beta_2 U$ where $\beta_1, \beta_2$ are the mechanical coupling coefficients. In system~\eqref{EQS} either $J_T$ or $J_X$ are used as coupling forces and $F_3 (Z,J,U)$ follows expressions \eqref{Fkolm}.

\subsection*{Initial and boundary conditions}
A $\sech^{2}$-type localized initial condition with an initial amplitudes $Z_o$ and $J_o$ are applied to $Z$ and $J$ in system~\eqref{EQS} and we make use of the periodic boundary conditions for all the members of the model equations
\begin{equation} \label{algtingimus}
\begin{split}
& Z(X,0) = Z_{o} \sech^2 B_{o} X, \quad Z(X,T) = Z (X + 2 K m \pi,T), \quad m = 1,2,\ldots ,\\
& J(X,0) = 0, \quad J_T(X,0) = 0, \quad J(X,T) = J (X + 2 K m \pi,T), \quad m = 1,2,\ldots ,\\
& U(X,0) = 0, \quad U_T(X,0) = 0, \quad U(X,T) = U (X + 2 K m \pi,T), \quad m = 1,2,\ldots ,\\
& \bar{P}(X,0) = 0, \quad \bar{P}_T(X,0) = 0, \quad \bar{P}(X,T) = \bar{P} (X + 2 K m \pi,T), \quad m = 1,2,\ldots ,\\
& \Theta(X,0) = 0, \quad \Theta(X,T) = \Theta (X + 2 K m \pi,T), \quad m = 1,2,\ldots ,
\end{split}
\end{equation}
where $K$ is the total  number  $2\pi$ sections in the spatial period. The amplitude of the initial `spark' is $Z_o$ and the width parameter is taken as $B_o$. In a nutshell -- such an initial condition is a narrow `spark' in the middle of the considered space domain with the amplitude above the threshold resulting in the usual FHN action potential formation which then proceeds to propagate in the positive and negative directions of the 1D space domain under consideration. 
For all other equations we take initial excitation to be zero and make use of the same periodic boundary conditions. The solution representing the mechanical and pressure wave is generated over time as a result of coupling with the action potential and ion current parts in the model system. It should be noted that in the present paper wave interactions are not investigated and integration intervals in time are picked such that the waves modeled do not reach the boundaries so the type of boundary conditions used is of low importance. For making use of the pseudospectral method periodic boundary conditions are needed.

While not shown in the present paper it should be added that the action potentials (and ion currents tied to these) annihilate each other during the interaction (as expected) but the mechanical and pressure waves can keep on going through many interactions if one uses the fact that we have periodic boundary conditions for taking a look at the interactions of the modeled wave ensembles.

\subsection*{The derivatives and integration}
For numerical solving of the system~\eqref{EQS} the discrete Fourier transform (DFT) based (PSM) (see \cite{Fornberg1998,salup2009}) is used.
Variable $Z$ can be represented in the Fourier space as
\begin{equation} \label{dft}
\widehat{Z}(k,T) = \mathrm{F} \left[ Z \right]= \sum^{n-1}_{j=0}{Z(j \Delta X, T) \exp{\left(-\frac{2 \pi \mathrm{i} j k}{n} \right)}},
\end{equation}
where $n$ is the number of space-grid points ($n=2^{12}$ in the present paper), $\Delta X=2 \pi/n$ is the space step, $k=0,\pm1,\pm2,\ldots,\pm(n/2-1),-n/2$; $\mathrm{i}$ is the imaginary unit, $\mathrm{F}$ denotes the DFT and $\mathrm{F}^{-1}$ denotes the inverse DFT.
The idea of the PSM is to approximate space derivatives by making use of the DFT
\begin{equation} \label{dft2}
\frac{\partial^{m} Z}{\partial X^{m}} = \mathrm{F}^{-1}\left[(\mathrm{i} k)^{m} \mathrm{F}(Z) \right],
\end{equation}
reducing therefore the partial differential equation (PDE) to an ordinary differential equation (ODE) and then to use standard ODE solvers for integration with respect to time. The model~\eqref{EQS} contains a mixed derivative term and coupling force terms can be taken either as a space derivative which can be found like in Eq.~\eqref{dft} or time derivative which is not suitable for a direct PSM application and need to be handled separately.

For integration in time the model system~\eqref{EQS} is rewritten as a system of first order ODE's  after the modification to handle the mixed partial derivative term and a standard numerical integrator is applied. In the present paper ODEPACK FORTRAN code (see \cite{ODE}) ODE solver is used by making use of the F2PY (see \cite{F2PY}) generated Python interface. Handling of the data and initilization of the variables is done in Python by making use of the package SciPy (see \cite{SciPy}).

\subsection*{Handling of mixed derivatives}
Normally the PSM algorithm is intended for $ u_t = \Phi(u,u_x, u_{2x},\ldots,u_{mx})$ type equations. However, we have a mixed partial derivative term $H_2 U_{XXTT}$ in Eqs~\eqref{EQS} and as a result some modifications are needed (see \cite{lauriandrus2009,lauriandruspearu2007,salup2009}). 
Rewriting system~\eqref{EQS} the equation for $U$ so that all partial derivatives with respect to time are in the left-hand side of the equation
\begin{equation} \label{LHSofHE}
U_{TT} - H_2 U_{XXTT}= c^{2} U_{XX} + N U U_{XX} + M U^{2} U_{XX} + N \left( U_{X} \right)^2 + 2 M U \left(U_X \right)^2 - H_1 U_{XXXX}  + \gamma_1 {P}_T + \gamma_2 J_T
\end{equation}
allows one to introduce a new variable
$\Phi = U - H_2 U_{XX}.$
After that, making use of properties of the DFT, one can express the variable $U$ and its spatial derivatives in terms of the new variable $\Phi$:
\begin{equation}\label{UUXPhi}
U=\mathrm{F}^{-1}\left[\frac{\mathrm{F}(\Phi)}{1+H_2 k^2}\right],
\qquad
\frac{\partial^m U}{\partial X^m}
=\mathrm{F}^{-1}\left[\frac{(\mathrm{i} k)^m \mathrm{F}(\Phi)}{1+H_2 k^2}\right].
\end{equation}
Finally, in system~\eqref{EQS} the equation for $U$ can be rewritten in terms of the variable $\Phi$ as
\begin{equation} \label{HEtegelik}
\Phi_{TT} =
c^2 U_{XX} + N U U_{XX} + M U^{2} U_{XX} + N \left( U_{X} \right)^2 + 2 M U \left(U_X \right)^2 - H_1 U_{XXXX}  + \gamma_1 {P}_T + \gamma_2 J_T,
\end{equation}
where all partial derivatives of $U$ with respect to $X$ are calculated in terms of $\Phi$ by using  expression \eqref{UUXPhi} and therefore one can apply the PSM for numerical integration of Eq.~\eqref{HEtegelik}. Other equations in the  model \eqref{EQS} are already written in the form which can be solved by the standard PSM.

\subsection*{The time derivatives ${P}_T$ and $J_T$}
The time derivatives ${P}_T$ and $J_T$ are found using different methods. For finding ${P}_T$ it is enough to write the equation for ${P}$ in system~\eqref{EQS} as two first order ODE's which is done anyway as the integrator requires first order ODE's and it is possible to extract ${P}_T$ from there directly
\begin{equation}
\begin{split}
& {P}_{T} = \bar{V} \\
& \bar{V}_{T} = c_{f}^{2} {P}_{XX} + \eta_1 Z_X + \eta_2 J_T - \mu {P}_T.
\end{split}
\label{EQSp}
\end{equation}
For finding $J_T$ a basic backward difference scheme is used
\begin{equation}
J_T (n,T) = \frac{J (n,T) - J (n,(T-dT))}{T - (T - dT)} \approx \frac{\Delta J(n,T)}{d T},
\label{backdiff}
\end{equation}
where $J$ is the ion current from Eqs~\eqref{EQS}, $n$ is the spatial node number, $T$ is the dimensionless time and $dT$ is the integrator internal time step value (which is variable and in the present paper the integrator is allowed to take up to $10^6$ internal time steps between $\Delta T$ values to provide the desired numerical accuracy. See technical details section for the notes on handling first time step and the last integrator internal time step (where $dT=0$).

\subsection*{The technical details and numerical accuracy}
As noted, the calculations are carried out with the Python package SciPy (see \cite{SciPy}), using the FFTW library (see \cite{FFTW3}) for the DFT and the F2PY (see \cite{F2PY}) generated Python interface to the ODEPACK FORTRAN code (see \cite{ODE}) for the ODE solver. The particular integrator used is the  `vode' with options set to nsteps$=10^6$, rtol$=1e^{-11}$, atol$=1e^{-12}$ and $\Delta T = 2$.

It should be noted that typically the hyperbolic functions like the hyperbolic secant $\sech^2(X)$ in our initial conditions in \eqref{algtingimus} are defined around zero. However, in the present paper the spatial period is taken from $0$ to $K \cdot 2\pi$ which means that the noted functions in \eqref{algtingimus} are actually shifted to the right (in direction of the positive axis of space) by $K \cdot \pi$ so the shape typically defined around zero is actually in our case located in the middle of the spatial period. This is a matter of preference (in the present case the reason is to have more convenient mapping between the values of $X$ and indexes) and the numerical results would be the same if one would be using a spatial period from $-K \cdot \pi$ to $K \cdot \pi$.

The `discrete frequency function' $k$ in \eqref{dft2} is typically formulated on the interval from $-\pi$ to $\pi$, however, we use a different spatial period than $2\pi$ and also shift our space to be from $0$ to $K \cdot 2\pi$ meaning that
\begin{equation}
k = \left[\frac{0}{K}, \frac{1}{K}, \frac{2}{K}, \ldots, \frac{n/2 - 1}{K}, \frac{n/2}{K}, - \frac{n/2}{K}, - \frac{n/2 - 1}{K}, \ldots , - \frac{n - 1}{K}, - \frac{n}{K} \right],
\label{diskreetnesagedus}
\end{equation}
where $n$ is number of the spatial grid points uniformly distributed across our spatial period (the size of the Fourier spectrum is $(n/2)$ which is, in essence, the number of spectral harmonics used for approximating the periodic functions and their derivatives) and $K$ is the number of $2\pi$ sections in our space interval.

There is few different possibilities for handling the division by zero rising in Eq.~\eqref{backdiff} during the initial initialization of the ODE solver and when the numerical iteration during the integration reaches the desired accuracy resulting in a zero length time step. For initial initialization of the numerical function initial value of $1$ is used for $dT$. This is just a technical nuance as during the initialization the time derivative will be zero anyway as there is no change in the value of $J(n,0)$. For handing the division by zero during the integration when ODE solver reaches the desired accuracy using values from two steps back from the present time for $J$ and $T$ is computationally the most efficient. Another straightforward alternative is using a logical cycle inside the ODE solver for checking if $dT$ would be zero but this is computationally inefficient. In the present paper using a value two steps back in time for calculating $J_T$ is used for all presented results involving $J_T$. The difference between the numerical solutions of the $J_T$ with the scheme using a value 1 step back and additional logic cycle for cheeking for division by zero and using two steps back in time scheme only if division by zero occurs is only approximately $10^{-6}$ and is not worth the nearly twofold increase in the numerical integration time.

Overall accuracy of the numerical solutions is approximately $10^{-7}$ for the fourth derivatives, approximately $10^{-9}$ for the second derivatives and approximately $10^{-11}$ for the time integrals. The accuracy of $J_T$ is approximately $10^{-6}$ which is adequate and very roughly in the same order of magnitude as the fourth spatial derivatives. The accuracy estimates are not based on the solving system~\eqref{EQS} with the presented parameters and are based instead on using the same scheme with the same technical parameters for finding the derivatives of $\mathrm{sin}(x)$ and comparing these to an analytic solution. In addition it should be noted that in the PST the spectral filtering is a common approach for increasing the stability of the scheme -- in the numerical simulations for the present paper the filtering (suppression of the higher harmonics in the Fourier spectrum) is not used although the highest harmonic (which tends to collect the truncation errors from the finite numerical accuracy of floating point numbers in the PST schemes) is monitored as a `sanity check' of the scheme.

\subparagraph{Acknowledgments.}
This research was supported by the European Union through the European Regional Development Fund (Estonian Programme TK 124) and presently KT and TP have support from the Estonian Research Council (projects IUT 33-24, PUT 434).

%


\addcontentsline{toc}{section}{References}

\end{document}